\documentclass[12pt]{article}
\usepackage[english]{babel}
\usepackage{amsmath}
\usepackage{amsthm}
\usepackage{amsfonts}
\usepackage{amssymb}
\usepackage{euscript}
\usepackage[cp1251]{inputenc}
\usepackage[pdftex]{graphicx}
\usepackage[12pt]{extsizes}

\begin{document}
\begin{center}
  \textbf{Common eigenfunctions of commuting\\ differential operators of rank 2}
\end{center}
\begin{center}
  \textbf{V.Oganesyan}
\end{center}
\begin{center}
   \textbf{Introduction}
\end{center}
Let us consider two differential operators
\begin{equation*}
L_n= \sum\limits^{n}_{i=0} u_i(x)\partial_x^i,  \quad  L_m= \sum\limits^{m}_{i=0} v_i(x)\partial_x^i.
\end{equation*}
If $L_n$ and $L_m$ commute, then there is a nonzero polynomial $R(z,w)$ such that  $R(L_n,L_m)=0$ (see ~\cite{Chaundy}). The curve $\Gamma$ defined by $R(z,w)=0$ is called the \emph{spectral curve}. If\\
\begin{equation*}
L_n \psi=z\psi, \quad  L_m \psi=w\psi,
\end{equation*}
then $(z,w) \in \Gamma$. For almost all $(z,w) \in \Gamma$ the dimension of the space of common eigenfunctions $\psi$ is the same. The dimension of the space of common eigenfunctions of two commuting differential operators is called the \emph{rank}. The rank is a common divisor of m and n.\\
If the rank equals 1, then there are explicit formulas for coefficients of commutative operators in terms of Riemann theta-functions (see ~\cite{theta}).\\
The case when rank is greater than one is much more difficult. The first examples of commuting ordinary scalar differential operators of the nontrivial ranks 2 and 3 and the nontrivial genus g=1 were constructed by Dixmier ~\cite{Dixmier} for the nonsingular elliptic spectral curve $w^2=z^3-\alpha$, where $\alpha$ is arbitrary nonzero constant:
\begin{equation*}
\begin{gathered}
L= (\partial_x^2 + x^3 + \alpha)^2 + 2x ,\\
M= (\partial_x^2 + x^3 + \alpha)^3 + 3x\partial_x^2 + 3\partial_x + 3x(x^2+\alpha) ,
\end{gathered}
\end{equation*}
where $L$ and $M$ is the commuting pair of the Dixmier operators of rank 2, genus 1. There is an example
\begin{equation*}
\begin{gathered}
L= (\partial_x^3 + x^2 + \alpha)^2 + 2\partial_x  ,\\
M= (\partial_x^3 + x^2 + \alpha)^3 + 3\partial_x^4 +  3(x^2+\alpha)\partial_x + 3x  ,
\end{gathered}
\end{equation*}
where $L$ and $M$ is the commuting pair of the Dixmier operators of rank 3, genus 1.\\
The general classification of commuting ordinary differential operators of rank greater than 1 was obtained by Krichever ~\cite{ringkrichever}. The general form of commuting operators of rank 2 for an arbitrary elliptic spectral curve was found by Krichever and Novikov ~\cite{novkrich}. The general form of operators of rank 3 for an arbitrary elliptic spectral curve was found by Mokhov ~\cite{Mokhov1},~\cite{Mokhov2}.\\
Mironov in ~\cite{Mironov} constructed examples of operators
\begin{equation*}
\begin{gathered}
L = (\partial_x^2 + A_3x^3+ A_2x^2 + A_1x + A_0 )^2 + g(g+1)A_3x ,\\
M^2 = L^{2g+1} + a_{2g}L^{2g} + ... + a_1L + a_0 ,
\end{gathered}
\end{equation*}
where $a_i$  are some constants and $A_i$, $A_3 \neq 0$, are arbitrary constants. Operators $L$ and $M$ are commuting operators of rank 2, genus g.\\
Examples of commuting ordinary differential operators of arbitrary genus and arbitrary rank with polynomial coefficients were constructed in  ~\cite{Mokhov4} by Mokhov.\\
It is proved in ~\cite{Vartan} that the operators
\begin{equation*}
\begin{gathered}
L=(\partial_x^2 + Ax^6 + Bx^2 )^2 + 16g(g+1)Ax^4, \\
M^2=L^{2g+1} + a_{2g}L^{2g} + ... + a_1L + a_0,
\end{gathered}
\end{equation*}
where $A, B$ are arbitrary constants, $A\neq 0$, $a_i$ are some constants, are commuting operators of rank 2.\\
In this paper we find common eigenfunctions of $L$ and $M$. Until now common eigenfunctions of commuting differential operators with polynomial coefficient were not found explicitly. \\
\\
The author is grateful to O.I.Mokhov for valuable discussions.
\begin{center}
   \textbf{Commuting operators of rank 2}
\end{center}
Consider the operator
\begin{equation*}
L=(\partial^2_x + V(x))^2 + W(x).
\end{equation*}
 We know that (\cite{Mironov})  the operator commutes with an operator M of order $4g+2$ with hyperlyptic spectral curve of genus $g$ and hence is operator of rank 2, if and only if there is a polynomial
\begin{equation*}
Q=z^g + a_1(x)z^{g-1} + a_2(x)z^{g-2} + ...+a_{g-1}(x)z + a_g(x)
\end{equation*}
that the following relation is satisfied
\begin{equation*}
Q^{(5)} + 4VQ''' + 6V'Q'' + 2Q'(2z-2W+V'') - 2QW'\equiv 0,
\end{equation*}
where $Q'$ means $\partial_xQ$. The spectral curve has the form
\begin{equation*}
4w^2=4(z-W)Q^2 - 4V(Q')^2 + (Q'')^2 - 2Q'Q''' + 2Q(2V'Q' + 4VQ'' + Q^{(4)}).
\end{equation*}
Common eigenfunctions of $L$ and $M$ satisfy the second order differential equation ~\cite{Mironov}
\begin{equation*}
\psi''(x,P) - \chi_{1}(x,P) \psi'(x,P) - \chi_0(x,P) \psi(x,P) =0,
\end{equation*}
where  $\chi_0$ and $\chi_1$ have the form
\begin{equation*}
\chi_1 = \frac{Q'}{Q}, \quad \chi_0 = -\frac{Q''}{2Q} + \frac{w}{Q} - V.
\end{equation*}
\begin{center}
   \textbf{Common eigenfunctions of commuting differential operators of rank 2}
\end{center}
Let me recall some definitions.\\
Bessel functions $J_{\alpha}$ are solutions of the Bessel equation
\begin{equation*}
x^2y''  + xy' + (x^2 - \alpha^2)y=0.
\end{equation*}
If $\alpha$ is not integer, then $J_{\alpha}$, $J_{-\alpha}$ satisfy Bessel equation, where
\begin{equation*}
J_{\alpha}(x)=\frac{x^{\alpha}}{2^{\alpha} \Gamma (\alpha+1)}(1-\frac{x^2}{2^2 1!(\alpha+1) } + \frac{x^4}{2^4 2! (\alpha+2)}-...)
\end{equation*}
If $\alpha$ is integer, then $J_{\alpha}$, $J_{-\alpha}$ are not independent solutions. Note that (see \cite{Ince})
\begin{equation*}
J'_{\alpha} = \frac{\alpha J_\alpha(x)}{x} - J_{\alpha + 1}(x).
\end{equation*}
Functions
\begin{equation*}
Y_{\alpha}(x) = \frac{J_{\alpha}(x)cos(\alpha\pi) - J_{-\alpha}(x)}{sin(\alpha\pi)}
\end{equation*}
are called Bessel functions of the second kind.\\
Solutions $H(\alpha, \lambda, \beta,\gamma,\delta,\eta; x)$ of the following equation
\begin{equation*}
y''(x)  + (\frac{\gamma}{x} + \frac{\delta}{x-1} + \frac{\alpha + \beta - \gamma - \delta + 1}{x-a})y'(x) + \frac{\alpha \beta x - q}{x(x-1)(x-a)}y(x)=0.
\end{equation*}
are called Heun functions. This equation has four regular singular points $0,1,a,\infty$. Confluent Heun equation is obtained from the Heun equation through a confluence process (see ~\cite{Lay}), that is, a process where two singularities coalesce. Denote by $CH(\alpha, \beta,\gamma,\delta,\eta; x)$ the solution of the confluent Heun equation
\begin{equation*}
\begin{gathered}
y''(x)  + (\frac{\beta + \gamma - \alpha + 2}{x-1} + \frac{x\alpha}{x-1} - \frac{\beta + 1}{x(x-1)})y'(x) +\\
+(\frac{\alpha(\beta + \gamma + 2) + 2\delta}{2(x-1)} -\frac{\alpha(\beta+1) - \beta(\gamma+1) - 2\eta - \gamma)}{2x(x-1)})y(x)=0
\end{gathered}
\end{equation*}
where
\begin{equation*}
\begin{gathered}
y(0)=1$,\quad $y'(0)=\frac{\beta(\gamma-\alpha+1) + \gamma - \alpha + 2\eta}{2(\beta+1)}.
\end{gathered}
\end{equation*}
There is formula for Bessel functions ~\cite{Lay}
\begin{equation*}
J_{\alpha}(x)=\frac{x^{\alpha}(2ix + 1) CH(1,2\alpha,1,0,\frac{1}{2}; -2ix)}{\Gamma(\alpha+1) 2^{\alpha} e^{ix}}.
\end{equation*}
We know from ~\cite{Vartan} that
\begin{equation*}
L=(\partial_x^2 + Ax^6 + Bx^4)^2 + 16g(g+1)Ax^4
\end{equation*}
commutes with a differential operator M of order $4g+2$. Let us assume that $B=0$. If $g=1$, then spectral curve of commuting pair L and M is equal to
\begin{equation*}
w^2=z(192A + z^2)
\end{equation*}
and differential equation for common eigenfunctions has the form
\begin{equation}
\psi'' -\frac{64Ax^3}{16Ax^4 + z} \psi' - (\frac{w - 96Ax^2}{16Ax^4 + z} - Ax^6)\psi=0.
\end{equation}
Let us suppose that $w=0$. So,  $z=0, \pm\sqrt{-192A}$. If $z=0$, then solutions of (1) are
\begin{equation*}
x^{\frac{5}{2}} J_{\frac{1}{8}}(\frac{x^4 \sqrt{A}}{4}), \quad x^{\frac{5}{2}} Y_{\frac{1}{8}}(\frac{x^4 \sqrt{A}}{4}).
\end{equation*}
If $z=\pm\sqrt{-192A}$, then solutions of (1) are
\begin{equation*}
\begin{gathered}
e^{-\frac{Ax^4}{4}\sqrt{-\frac{1}{A}}} CH(\frac{z}{32}\sqrt{-\frac{1}{A}},  -\frac{1}{4}, -2 , 0, \frac{5}{4}; -\frac{16Ax^4}z), \\
xe^{-\frac{Ax^4}{4}\sqrt{-\frac{1}{A}}} CH(\frac{z}{32}\sqrt{-\frac{1}{A}},  \frac{1}{4}, -2 , 0, \frac{5}{4}; -\frac{16Ax^4}z).
\end{gathered}
\end{equation*}
If $g=2$, then spectral curve of commuting operators L and M is equal to
\begin{equation*}
w^2 = z(20160A + z^2)(20736A + z^2).
\end{equation*}
If $z=0$, then equation for common eigenfunctions has the form
\begin{equation}
(4Ax^8 + 35)\psi'' -32Ax^7\psi' +(147Ax^6 + 4A^2x^{14})\psi=0.
\end{equation}
Equation (2) has solutions
\begin{equation*}
\begin{gathered}
CH(0,-\frac{1}{8},-2,-\frac{35}{256}, \frac{387}{256}; -\frac{4Ax^8}{35}),\\
xCH(0,\frac{1}{8},-2,-\frac{35}{256}, \frac{387}{256}; -\frac{4Ax^8}{35}).
\end{gathered}
\end{equation*}

Department of Geometry and Topology, Faculty of Mechanics and Mathematics, Lomonosov Moscow State University, Moscow, 119991 Russia.\\\\
E-mail address: vardan.o@mail.ru


\begin{thebibliography}{10}

\bibitem{Chaundy}Burchnall J.-L., Chaundy T.W. Commutative ordinary differential operators.
Proc. London Math. Soc. 21 (1923), 420-440; Proc. Royal Soc. London (A)
118 (1928), 557-583.
\bibitem{theta}I.M. Krichever, Integration of nonlinear equations by the methods of algebraic geometry, Func-
tional Anal. Appl., 11: 1 (1977), 12–26.
\bibitem{ringkrichever}I.M. Krichever. Commutative rings of ordinary linear differential operators.Funktsional. Anal. i Prilozhen.,12:3 (1978), 20–31 (in Russian);English translation inFunctional Anal. Appl.,12:3 (1978), 175–185.
\bibitem{novkrich}I.M. Krichever, S.P. Novikov. Holomorphic bundles and nonlinear equations. Finite-zone solutions of rank 2.Dokl. Akad. Nauk SSSR,247:1 (1979), 33–36(in Russian);English translation in Soviet Math. Dokl..
\bibitem{Mokhov1}O.I. Mokhov. Commuting ordinary differential operators of rank 3 corresponding to an elliptic curve. Uspekhi Matem. Nauk,37:4 (1982), 169–170 (in Russian); English translation in Russian Math. Surveys,37:4 (1982), 129–130.
\bibitem{Mokhov2}O.I. Mokhov. Commuting differential operators of rank 3, and nonlinear differential equations.Izvestiya AN SSSR, ser. matem.,53:6 (1989), 1291–1315 (in Russian); English translation in Math. USSR, Izvestiya,35:3 (1990), 629–655.
\bibitem{Mironov}A.E. Mironov. Self-adjoint commuting differential operators and commutative subalgebras of the Weyl algebra. Invent. math. (2014) 197:417-431
\bibitem{Dixmier} J. Dixmier,Bull. Soc. Math. France,96(1968), 209–24
\bibitem{Mironov2} A.E.Mironov. Periodic and rapid decay rank two self-adjoint commuting differential operators arXiv:1302.5735
\bibitem{Mokhov3} O. I. Mokhov. On Commutative Subalgebras of the Weyl Algebra Related to Commuting Operators of Arbitrary Rank and Genus”, Mat. Zametki, 94:2 (2013), 314–316 (in Russian); English translation in Mathematical Notes, vol. 94:2 (2013), 298-300
\bibitem{Mokhov4}  O.I.Mokhov. Commuting ordinary differential operators of arbitrary genus and arbitrary rank with polynomial coefficients. American Mathematical Society Translations, Volume 234 (2014),  323-336.
\bibitem{Vartan}  V.S.Oganesyan. Commuting differential operators of rank 2 with polynomial coefficients.  arXiv:1409.4058
\bibitem{Ince}  E.L.Ince. Ordinary differential equations, Dover, New York, 1956.
\bibitem{Lay}  S. Y. Slavyanov, W. Lay. Special Functions: A Unified Theory Based on Singularities.

\end{thebibliography}
\end{document}